\documentclass[a4paper,11pt]{article}
\pdfoutput=1 
\usepackage{jheppub} 

\usepackage[utf8]{inputenc}
\usepackage{graphicx,rotating,amsmath,amsfonts,amssymb}

\usepackage{epsfig}
\usepackage{hyperref}
\usepackage{color}

\newcommand{\ie}{{\it i.e. }}
\newcommand{\eg}{{\it e.g. }}

\newcommand{\be}{\begin{equation}}
\newcommand{\ee}{\end{equation}}
\newcommand{\br}{\begin{eqnarray}}
\newcommand{\bea}{\begin{eqnarray}}
\newcommand{\eea}{\end{eqnarray}}
\newcommand{\er}{\end{eqnarray}}
\newcommand{\ba}{\begin{array}}
\newcommand{\ea}{\end{array}}
\newcommand{\bi}{\begin{itemize}}
\newcommand{\ei}{\end{itemize}}
\newcommand{\bn}{\begin{enumerate}}
\newcommand{\en}{\end{enumerate}}
\newcommand{\bc}{\begin{center}}
\newcommand{\ec}{\end{center}}

\newcommand{\bra}[1]{\langle #1 \vert}
\newcommand{\ket}[1]{\vert #1 \rangle}
\newcommand{\braket}[2]{\langle #1 \vert #2\rangle}

\newcommand{\td}{\mathrm{d}}

\newcommand{\conj}{{\rm \star}}

\def\gappeq{\mathrel{\rlap {\raise.5ex\hbox{$>$}}
{\lower.5ex\hbox{$\sim$}}}}

\def\lappeq{\mathrel{\rlap{\raise.5ex\hbox{$<$}}
{\lower.5ex\hbox{$\sim$}}}}

\title{On the Quantisation of Complex  Higher Derivative Theories and Avoiding  the Ostrogradsky Ghost}

\author[a]{Martti Raidal,}
\author[a]{and Hardi Veerm\"ae}

\emailAdd{martti.raidal@cern.ch}
\emailAdd{hardi.veermae@cern.ch}

\affiliation[a]{National Institute of Chemical Physics and Biophysics,  R\"avala 10, 10143 Tallinn, Estonia}

\abstract{Generic higher derivative theories are believed to be fundamentally unphysical because they contain Ostrogradsky ghosts. 
We show that within complex classical mechanics it is possible to construct higher derivative theories that circumvent the Ostrogradsky theorem and have a real energy spectrum that is bounded from below. The complex theory can be canonically quantised. The resulting quantum theory does not suffer from the kinetic instability and maintains the usual probabilistic interpretation without violating the correspondence principle.  As a proof of concept, we construct a class of stable interacting complex higher derivative theories and present a concrete example. This consistent and canonical framework allows us to analyse the previous attempts to avoid ghosts that use non-canonical quantisation schemes, such as the Lee-Wick theories, Dirac-Pauli quantisation or $PT$-symmetric quantum mechanics. The key to understand the would-be ghosts in any kinetically stable higher derivative theory is to accept the complex system behind it.}

\begin{document}

\maketitle
\flushbottom

\section{Introduction}
\label{1}

The theorem by Ostrogradsky~\cite{Ostrogradsky:1850fid} formulated more than one and a half centuries ago is likely the reason why all fundamental equations of motion of physical theories known to date contain at most second time derivatives. In the presence of higher derivatives it predicts unbounded kinetic terms that inevitably lead to runaway solutions if interactions are turned on. The Ostrogradsky theorem is thus a deeply fundamental result comparable to the symmetry principles in modern quantum field theory. By general arguments its pathological consequences are present both in classical as well as in quantum theories~\cite{Woodard:2015zca}.

On the other hand, some quantum field theories may actually benefit form the existence of higher derivatives. The most important example would be the improved renormalizability present in Lee-Wick theories~\cite{Lee:1969fy,Lee:1970iw} or in higher derivative gravity~\cite{Stelle:1976gc, Modesto:2015ozb}. In the latter case it has been proven by Stelle that inclusion of possible terms quadratic in the Riemann tensor results in a renormalizable theory of quantum gravity. However, due to the fatal nature of the Ostrogradsky instability, this idea seems to have been born dead. 

Higher derivatives make frequent appearances through higher order operators of effective field theories. If these terms are treated perturbatively, as they should, the ghost degrees of freedom do not appear. Technically this is because the propagators will not develop additional poles in a perturbative expansion~\cite{Simon:1990ic}. The downside of this fact is that the extra poles were precisely the reason for the improved renormalizability in higher derivative theories. Thus, at first sight, it seems that we are given the choice between the non-renormalizability and the Ostrogradsky ghosts and, with a good reason, the former is commonly considered to be the lesser evil of the two.

Several attempts of curing the Ostrogradsky instability are based on modifying the quantisation scheme. For example, the indefinite metric Dirac-Pauli quantisation~\cite{Dirac:1942,Pauli:1943,Salvio:2015gsi} used to quantise the Lee-Wick theories~\cite{Lee:1969fy,Lee:1970iw}, their extensions~\cite{Grinstein:2007mp,Grinstein:2008bg,Grinstein:2008qq} or A-gravity~\cite{Salvio:2014soa} involve negative norm states. Secondly,  the $PT$-symmetric quantum mechanics~\cite{Bender:1998ke} involving non-Hermitian Hamitonians and multiple inner products has also been used to eliminate the ghosts~\cite{Bender:2007wu,Bender:2008gh,Bender:2007nj,Lee:2013wfa}. All these quantum theories predict a positive spectrum of the Hamiltonian and therefore, by general arguments, are at odds with the classical-to-quantum correspondence~\cite{Woodard:2015zca}. In that case it is difficult to tell which classical theory one is dealing with and, moreover, whether these quantum theories describe a higher derivative theory at all.

In this paper we take a canonical approach of circumventing the Ostrogradsky theorem already at the classical level. We achieve this by complexifying the higher derivative theories. These complex systems can be consistently quantised using the rules of canonical quantisation.  In order for the energy spectrum of the theory be bounded, we show that the would-be ghost degrees of freedom are necessarily complex.  As a result, one obtains a quantum theory that possesses all good properties of the known quantum physics including a clear classical-to-quantum correspondence, the standard probabilistic interpretation, and no Ostrogradsky instability. As a proof of concept we present a class of non-trivial interacting complex higher derivative theories that could, indeed, describe stable physical systems, and exemplify this with a concrete example model.

The quantum physics of the proposed framework is canonical and consistent. Therefore it allows us to analyse the previous attempts of exorcising the ghosts by using non-canonical quantisation schemes. Using the prototypical Pais-Uhlenbeck oscillator, we demonstrate that quantisation with negative norm states~\cite{Dirac:1942,Pauli:1943,Salvio:2015gsi}, for example Lee-Wick theories~\cite{Lee:1969fy,Lee:1970iw}  or A-gravity~\cite{Salvio:2014soa}, and also the $PT$-symmetric treatment~\cite{Bender:2007wu,Bender:2008gh,Bender:2007nj,Lee:2013wfa}, are equivalent to the canonically quantised complex Pais-Uhlenbeck oscillator with a real and an imaginary degree of freedom. The imaginary degree of freedom would correspond to the ghost in the real case, yet in the complex theory it carries positive energy. Thus we will refer to it as ``the imaginary ghost". While in the $PT$-symmetric treatment one can formally create a mapping between the non-Hermitian and the Hermitian one (although not always, as claimed in Ref.~\cite{Bender:2008gh}), we show that non-canonical quantisation unnecessarily complicates both the interpretation and the mathematical treatment of the theory. At the same time it provides little insight into the problem of the ghost. 

The most important result of this work is finding the consistent and canonical principle behind the interpretation of physically viable higher derivative degrees of freedom. If any higher derivative theory will turn out to describe Nature, for example gravity, then it must be complex for the spectrum to be bounded from below. Non-standard quantisation is not needed. Such tricks add little to the understanding of ghosts and, surprisingly, they seem to have confused physicists for the last 75 years. Our result of establishing the physical would-be ghosts as canonically quantised complex objects may set the future research of higher derivative theories on a well defined path. 

Whether the complex higher derivative theories will be of any use to describe physical reality remains an open question. The complexification is, however, likely to retain the properties needed for the renormalizability of higher derivative gravity~\cite{Stelle:1976gc}. This provides another important application of the considered framework.

For completeness we note that the Ostrogradsky instability might be avoided in higher derivative systems with constraints~\cite{Chen:2012au}. An interesting mechanical example is the relativistic point particle with rigidity~\cite{Plyushchay:1988wx,Plyushchay:1990cu}. Additionally, complex quantum mechanics has been studied in the path integral 
approach~\cite{Witten:2010zr}. We will, however, not consider these systems in this work.

This work is structured as follows. 
In section~\ref{2} we outline general properties of higher derivative theories.
In section~\ref{3} we propose the idea of complex higher derivative theories and compare it to the alternative non-canonical quantisation schemes.
We conclude in section~\ref{4}.
Some of the technical details are collected to appendix~\ref{app}.

\section{General properties of higher derivative theories}
\label{2}

Higher derivative theories have a set of general properties that are independent of whether the theory is quantum or classical. First, they can always be converted to equivalent ordinary theories by replacing higher derivatives with new degrees of freedom. Second, as a result of the Ostrogradsky theorem, the kinetic terms of these new degrees of freedom are not bounded from below. Instead of studying higher derivative theories we could likewise consider normal theories with unbounded kinetic terms. Moreover, in some cases it is possible to invert this construction and replace a kinetically unstable theory with an equivalent higher derivative one. In this section we review these general features.

\subsection{Higher derivatives versus new degrees of freedom}
\label{sec:2.1}

We refer to a theory as normal if its equations of motion contain at most second derivatives. This implies that the corresponding Lagrangian contains, up to a total derivative, at most first time derivatives of the dynamical variables (for a proof see appendix~\ref{app:der}). Otherwise we refer to it as a higher derivative theory or a higher derivative Lagrangian, respectively. This distinction is, however, somewhat arbitrary as any higher derivative Lagrangian can be reduced to a normal one by introducing new dynamical variables. This subsection reviews how this can be done in a systematic way by making use of Lagrange multipliers.

We focus on the Lagrangian formalism and limit the discussion mainly to one-dimensional second derivative theories of one variable. To avoid unnecessary complications with constraints we assume the Lagrangian $L(q,\dot{q}, \ddot{q})$ to be non-degenerate, \ie $\partial^{2} L / \partial^{2} \ddot{q} \neq 0$, but otherwise completely general. This Lagrangian can be reduced to a normal one by introducing the auxiliary variable $\lambda$ that is identified with the second derivative $\ddot{q}$ by using a Lagrange multiplier $Q$~\cite{Pons1989,Motohashi:2014opa}:
\begin{align}\label{def:Llambda}
	L(q,\dot{q}, \ddot{q}) \equiv L(q,\dot{q}, \lambda) + Q(\ddot{q} - \lambda).
\end{align}
The resulting system is constrained by construction. To remove the constraint we use the equations of motion of $\lambda$
\begin{align}\label{def:Q}
	Q \equiv \left.\frac{\partial L}{\partial \ddot{q}}\right|_{\ddot{q} = \lambda},
\end{align}
that can be solved if $\partial^{2} L / \partial^{2} \ddot{q}$ is either positive or negative definite, \ie if the initial Lagrangian is non-degenerate.
The solution yields an expression  $\lambda = \lambda(q,\dot{q}, Q)$. Finally, after eliminating the last remaining second derivative by partially integrating the $\ddot{q} Q$ term, we obtain an equivalent first derivative Lagrangian to \eqref{def:Llambda},
\begin{align}\label{def:Lprim}
	L'(q,\dot{q}, Q, \dot{Q}) = L(q,\dot{q}, \lambda(q,\dot{q}, Q)) - Q \lambda(q,\dot{q}, Q) - \dot{q} \dot{Q},
\end{align}
that we expressed explicitly as a function of the dynamical variables $q$, $Q$  and their first derivatives only. It is manifest that the theory describes two degrees of freedom. Its linearity in $\dot{Q}$ needs to be highlighted as it implies an unbounded kinetic term. The last statement, in essence, results in the Ostrogradsky instability.  

This procedure of trading a higher derivative with new a degree of freedom can be straightforwardly extended to higher dimensional spaces or to Lagrangians containing arbitrarily high derivatives. For example, take a $n$ derivative Lagrangian $L(q,\dot{q}, \ddot{q}, \ldots, q^{(n)})$ that is non-degenerate in $q^{(n)}$. It is equivalent to $L(q,\dot{q}, \ddot{q}, \ldots, \lambda)+ Q(q^{(n)} - \lambda)$. Eliminating $\lambda$ by using its equations of motion and partially integrating the highest derivative results in a manifestly $n-1$ derivative theory of $q$ and a new normal degree of freedom $Q$. We remark that eliminating $\lambda$ induces a Legendre transformation of the original Lagrangian. 

Unbounded kinetic terms appearing in a generic normal theory, on the other hand, do not necessarily indicate equivalence with a higher derivative theory. An exception occurs in the special case when a variable appears as or, to be more specific, can be made to appear as an auxiliary variable. As an example consider the one-dimensional field theory with a particle and a ghost with canonical kinetic terms,
\begin{align}\label{ex:L_0}
	L = \frac{1}{2} \dot{q}_{1}^{2} - \frac{1}{2}\dot{q}_{2}^{2} - V(q_{1},q_{2}).
\end{align}
Defining new variables  $q = (q_{1} + q_{2})/\sqrt{2}$ and $Q = (q_{1} - q_{2})/\sqrt{2}$ yields the equivalent Lagrangian
\begin{align}\label{ex:L_1}
	L = -Q\ddot{q} - V(q,Q),
\end{align}
where we removed all derivatives of $Q$ by partial integration. This Lagrangian can be promoted to a non-degenerate higher derivative one since the variable $Q$ is auxiliary and can thus be eliminated by using its equation of motion, $\ddot{q} = -\partial V / \partial Q$, if $\partial^{2} V / \partial Q^{2} \neq 0$. To make the example more specific  consider the potential 
\begin{align}
	V(q,Q) = f_{0}(q) + f_{1}(q) Q  + f_{2}(q) Q^{2}.
\end{align}
In that case we find that the second derivative equivalent to Eq.~\eqref{ex:L_1} is
\begin{align}\label{ex:L_2}
	L = \frac{(\ddot{q} + f_{1})^{2}}{4 f_{2}}  - f_{0}.
\end{align}
The two variables we started with have been replaced by a higher derivative one. Notably, the kinetic term is singular when $f_{2} = 0$, reflecting that the equivalency between \eqref{ex:L_1} and \eqref{ex:L_2} is not well defined in that case. This corresponds to the point where the equation $\ddot{q} = -\partial V / \partial Q$ is not invertible.

\subsection{The Ostrogradsky theorem in classical and quantum mechanics}
\label{sec:2.2}

The Ostrogradsky theorem states that the Hamiltonian resulting from a higher derivative Lagrangian is not bounded from below~\cite{Ostrogradsky:1850fid,Motohashi:2014opa}. In this section we outline a short proof for a pointlike particle both in the classical and quantum mechanics. The arguments presented here can be straightforwardly extended to systems with arbitrary number of degrees of freedom, to $n$-derivative theories, or to field theories. For a review see Ref.~\cite{Woodard:2015zca}.

We remark that sometimes it is believed that quantisation might cure the Ostrogradsky instability as it cured the classical instability of the hydrogen atom. This, however, is not the case and the kinetic instability always survives quantisation. Here we consider only canonical quantisation, \ie we assume that real classical variables correspond to Hermitian operators and satisfy canonical commutation relations derived from the Poisson brackets.

In the previous subsection  we demonstrated that any non-degenerate second derivative theory with one variable is equivalent  to a first order theory of two variables,
see Eq.~\eqref{def:Lprim}. We can therefore restrict our attention to general Lagrangians of the form 
 \begin{align}\label{eq:L_g_ost}
	L'(q,\dot{q}, Q, \dot{Q}) = L_{0}(q,\dot{q}, Q) - \dot{q} \dot{Q}.
\end{align}
For our purposes the only relevant property of $L_{0}$ is that it does not depend on $\dot{Q}$, or in other words, that the full Lagrangian depends linearly on $\dot{Q}$. 
The Hamiltonian corresponding to \eqref{eq:L_g_ost} has the general form
\begin{align}\label{H_second_der2}
	H(q,p,Q,P) =  H_{0}(q,Q,P) - Pp,
\end{align}
where the conjugate momenta are defined by $p \equiv \partial L'/\partial \dot{q}$ and $P \equiv \partial L'/\partial \dot{Q}$. For completeness we determine the Hamiltonian corresponding to Eq.~\eqref{def:Lprim} explicitly in Appendix~\ref{app:ostr} and show that it is related to Ostrogradsky's construction by the canonical transformation $Q \to -P$, $P \to Q$. Leaving the specifics aside, the most relevant property of the Hamiltonian \eqref{H_second_der2} is its linearity in $p$ that, as such, renders the Hamiltonian unbounded. This, and the fact that the Hamiltonian is the conserved quantity corresponding to the invariance under time translations, establishes the Ostrogradsky instability in the classical case.

It is left to show that also the quantum Hamiltonian is not bounded. We do this by explicitly constructing states with arbitrary energy expectation values. Canonical quantisation promotes the canonical variables to Hermitian operators that satisfy canonical commutation relations, $[q,p] = i = [Q,P]$.  We then pick a state $\ket{\psi}$ with a finite non-vanishing expectation value of $P$ and a unit norm, \ie  $\bra{\psi}P\ket{\psi} \neq 0$ and $\braket{\psi}{\psi} = 1$. From this state we construct the boosted state $\ket{\psi_{p_{0}}}$ by applying the unitary operator $\exp(-i q p_{0})$,
\begin{align}
	\ket{\psi_{p_{0}}} \equiv \exp(-i q p_{0})\ket{\psi}.
\end{align}
From the fact that $[q,H_{0}] = 0 = [q,P]$, it follows that the energy expectation value transforms as
\begin{align}
	\bra{\psi}H\ket{\psi} \to
	\bra{\psi_{p_{0}}}H\ket{\psi_{p_{0}}}
&	= \bra{\psi}H\ket{\psi} - p_{0}\bra{\psi}P\ket{\psi}.
\end{align}
It is therefore possible to obtain arbitrary energy expectation values by choosing $p_{0}$ appropriately. To summarise, in the quantised theory it is possible to arbitrarily boost the momentum $p$ and, as in the classical case, this results in a shift in energy \eqref{H_second_der2}. 

Finally we remark that a system with unbounded Hamiltonian is not necessarily unstable, if the positive and negative energy modes are not interacting. The instability is turned on by interactions. This can be easily understood in a thermal setting as any thermal fluctuations will drive the system to fill the entire available phase space which, as the energy is unbounded, is now infinite. The situation is even more dire in the quantum theory where the system will tunnel to lower energy states. The infinite volume of the available phase space in multidimensional theories causes any state to decay instantaneously.

\section{Complex higher derivative theories}
\label{3}

Motivated by the result that the Ostrogradsky instability in higher derivative theories cannot be cured by quantisation, we show in the following 
that the assumptions of Ostrogradsky's theorem can be consistently circumvented by complexification of the theory. While this approach might look 
problematic when applied to the space-time variables, complex quantum field theories are among  the most commonly studied theories in high energy physics. 
Thus we argue that the complex ghost can actually be a physical degree of freedom in quantum field theories with higher derivatives.
To show that, we first study the quantisation of higher derivative theories following the example of the simplest of them, the Pais-Uhlenbeck (P-U) oscillator~\cite{Pais:1950za}. In the past decades the Ostrogradsky ghost of the Pais-Uhlenbeck oscillator has gotten considerable attention~\cite{Smilga:2005gb,Smilga:2008pr,Mannheim:2004qz,Bender:2008gh,Bender:2007wu,Pavsic:2013mja,Pavsic:2013noa,Pavsic:2016ykq,Woodard:2015zca}. The particular model is known to be free of physical instabilities because the positive and negative energy degrees of freedom are not interacting. Nevertheless, it captures the properties we are interested in and the results can be applied to a more general class of interacting models, as we will show in the following. In essence, we show that the negative sign of energy of the ghost degrees of freedom can be flipped by simply demanding the ghost to be imaginary. An analogous sign flip appears if the ghost state is quantised using negative norm states~\cite{Lee:1969fy,Mannheim:2004qz,Bender:2008gh,Bender:2007wu,Salvio:2015gsi}. We show that the two approaches are related and quantisation with negative norm states is equivalent to canonically quantising a theory with an imaginary ghost degree of freedom. The equivalence holds up to a redefinition of the norm.  Similarly, we show that the $PT$-symmetric approach to the Pais-Uhlenbeck oscillator can be readily reproduced by the canonical framework. 

\subsection{The complex Pais-Uhlenbeck oscillator}
\label{sec:CPU}

The Pais-Uhlenbeck oscillator is defined by the Lagrangian
\begin{align}\label{eq:L_PU}
	L	=	-\frac{1}{2}\ddot{q}^{2} + \frac{1}{2}(\omega_1^{2} + \omega_2^{2})\dot{q}^{2} - \frac{1}{2} \omega_1^{2}\omega_2^{2} q^{2},
\end{align}
where $\omega_{1}$, $\omega_{2}$ are real frequencies. We restrict the treatment to the non-degenerate case, $\omega_{1} \neq \omega_2$. The Lagrangian produces a fourth order equation of motion
\begin{align}\label{eq:eom_PU}
	q^{(4)} + (\omega_1^{2} + \omega_2^{2})q^{(2)} + \omega_2^{2}q = 0,
\end{align}
that, given $\omega_{1} \neq \omega_2$, is solved by a superposition of two independent harmonic oscillators
\begin{align}\label{eq:PU_sol}
	q(t) = A_{1} e^{-i\omega_1 t} + B_{1} e^{i\omega_1 t} + A_{2} e^{i\omega_2 t} + B_{2} e^{-i\omega_2 t}.
\end{align}
Reality of the solution would dictate that the constants of integration $A_{i}$, $B_{i}$ obey $A_{i} = B_{i}^{*}$. However, as the Pais-Uhlenbeck oscillator here is complex, $A_{i}$ and $B_{i}$ are completely independent complex variables. The solution does not display instabilities because the negative and positive energy modes are not interacting. Nevertheless, the energy (see \eg \eqref{E_second_der}) corresponding to the solution \eqref{eq:PU_sol},
\begin{align}\label{eq:E_PU}
	E = 2 (\omega_{2}^2 - \omega_{1}^2) (A_{1} B_{1} \omega_{1}^2 - A_{2} B_{2} \omega_{2}^2),
\end{align}
is clearly not bounded from below if the oscillator is real, \ie if $A_{i} = B_{i}^{*}$.

To give the theory a more familiar face we reduce the number of derivatives using the procedure described in Section~\ref{sec:2.1}. Proceeding with the Legendre transformation we identify $Q \equiv \frac{\partial L}{\partial \ddot{q}} = -\ddot{q}$ and obtain from Eq. \eqref{def:Lprim} the expression
\begin{align}\label{eq:Lp_PU}
	L'	=	\frac{1}{2}(\omega_1^{2} + \omega_2^{2})\dot{q}^{2} - \dot{q} \dot{Q} - \frac{1}{2}\omega_1^{2}\omega_2^{2}q^{2} + \frac{1}{2}Q^{2} .
\end{align}
The change of variables
\begin{align}\label{eq:cPU_qp12}
	q = \frac{q_{1} + q_{2} }{\sqrt{\omega_2^{2} - \omega_1^{2}}}, \qquad
	Q = \frac{\omega_1^{2} q_{1} + \omega_2^{2} q_{2}}{\sqrt{\omega_2^{2} - \omega_1^{2}}},
\end{align}
brings the Lagrangian into the canonical form
\begin{align}\label{eq:Lp2_PU}
	L'	=	\frac{1}{2}\left( \dot{q}_{1}^{2} - \omega_1^{2}q_{1}^{2} \right)- \frac{1}{2}\left( \dot{q}_{2}^{2} - \omega_2^{2}q_{2}^{2} \right),
\end{align}
which makes explicit that the model describes two harmonic oscillators of which one is a ghost.
The momenta conjugate to the coordinates $q_{i}$ are
\begin{align}\label{eq:cPU_p12}
	p_1 \equiv \frac{\partial L'}{\partial \dot{q}_{1}} = \dot{q}_{1},
	\qquad
	p_2  \equiv \frac{\partial L'}{\partial \dot{q}_{1}} = -\dot{q}_{2}.
\end{align}
Note the minus sign in the definition of the ghost momentum. The complex (classical) variables
\begin{align}\label{eq:cPU_abab}
	a_i &= \sqrt{\frac{\omega_{i}}{2}}\left( q_{i} + \frac{i}{\omega_{i}}p_{i} \right),
	\qquad
	b_i = \sqrt{\frac{\omega_{i}}{2}}\left(q_{i} - \frac{i}{\omega_{1}}p_{i}\right),
\end{align}
become the creation and annihilation operators after quantisation. Their Poisson brackets read
\begin{align}\label{eq:cPU_ab}
	[a_{i}, b_{j}]_{P} = -i [q_{i}, p_{j}]_{P}  = -i \delta_{ij}.
\end{align}
The variables $a_{i}$ and $b_{i}$ correspond directly to the modes \eqref{eq:PU_sol} found by solving the higher derivative  equations of motion \eqref{eq:eom_PU}, \ie  $a_{1} = A_{1} \omega_{1} \sqrt{ 2\omega_{1} (\omega_{2}^{2} - \omega_{1}^{2})} e^{- i\omega_1 t}$. Most of this is expected, with the minor exception that the frequency of the ghost modes $A_{2}$, $B_{2}$ in \eqref{eq:PU_sol} has an opposite signs compared to the normal modes. 

Quantisation proceeds by promoting the dynamical variables to operators and Poisson brackets to commutation relations by demanding that $[A, B] _{P} = c$ is replaced by  $[A, B]  = i c$. The Poisson brackets \eqref{eq:cPU_ab} thus imply
\begin{align}\label{eq:qPU_ab}
	[a_{i}, b_{i}]  = \delta_{ij}.
\end{align}
The operators $a_{i}$, $b_{i}$ are otherwise unrelated, especially $a_{i}^{\dagger} \neq b_{i}$, since the canonical coordinates are complex. The Hamiltonian  reads
\begin{align}\label{eq:cPU_H_a1a2}
	H = \frac{\omega_1}{2} \{a_{1},b_{1}\} - \frac{\omega_2}{2} \{a_{2},b_{2}\},
\end{align}
where the curly brackets denote anticommutators which impose a symmetric ordering of $a_{i}$, $b_{i}$. The commutation relations \eqref{eq:qPU_ab} together with the Heisenberg equations, $\dot{A} = i [H,A]$,  imply 
\begin{align}\label{eq:cPU_sol_ab}
	a_{1}(t) = a_{1} e^{-i \omega_1 t}, \qquad
	b_{1}(t) = b_{1} e^{i \omega_1 t}, \qquad
	a_{2}(t) = a_{2} e^{i \omega_2 t}, \qquad
	b_{2}(t) = b_{2} e^{-i \omega_2 t}.
\end{align}
Thus we conclude that the classical complex solution \eqref{eq:PU_sol} withstands canonical quantisation in the Heisenberg picture.

The interpretation of the obtained quantum theory is problematic because  the Hamiltonian \eqref{eq:cPU_H_a1a2} is not Hermitian by construction and the time evolution is therefore not unitary. Moreover, the solution \eqref{eq:cPU_sol_ab} is purely algebraic. If we would have chosen the Schr\"odinger equation as the starting point, then the operators would evolve according to $\dot{A} = i (H A - A H^{\dagger})$ and the classical-to-quantum correspondence would be lost. The resolution of this issue is beyond the scope of this paper and we merely remarked on some of mathematical properties of this theory. Especially that, from the algebraic point of view, canonical quantisation seems to agree with complex variables. Yet in the following we will mainly focus on special cases in which a Hermitian Hamiltonian might be constructed.

\subsection{The real and the imaginary ghost}
\label{sec:ReGImG}

Different possibilities exist for constraining the coordinates on the complex plane in order to make the Hamiltonian \eqref{eq:cPU_H_a1a2} or, equivalently, the energy \eqref{eq:E_PU} real. As the two modes are independent we need to constrain them independently if it is to be dynamically consistent.

First, the standard Pais-Uhlenbeck oscillator has real coordinates and therefore obeys
\begin{align}\label{eq:rPU_qp}
	q_i = \frac{1}{\sqrt{2\omega_{i}}}(a_{i} + b_{i}) = q_i^{\dagger},
	\qquad
	p_i  = \frac{\sqrt{\omega_{i}}}{i\sqrt{2}}(a_{i} - b_{i}) = p_i^{\dagger},
\end{align}
implying $b_{i} = a_{i}^{\dagger}$. The Hamiltonian \eqref{eq:cPU_H_a1a2} now reads
\begin{align}\label{eq:PU_H_a1a2}
	H = \frac{\omega_1}{2} \{a_{1},a^{\dagger}_{1}\} - \frac{\omega_2}{2} \{a_{2},a^{\dagger}_{2}\}.
\end{align}
It is clearly Hermitian. Consistency dictates that the ground state obeys
\begin{align}
	a_{1} \ket{0} = 0, \qquad a_{2} \ket{0} = 0,
\end{align}
while all other eigenstates can be obtained by the usual procedure of applying creation operators $a^{\dagger}_{i}$. As a result we obtain an unbounded spectrum. Despite this, the time evolution is still well defined and unitary -- the ghost cannot create problems, if it does not interact. Moreover, since the Hamiltonian remains Hermitian also in the presence of interactions, unitarity violation is unlikely if the corresponding classical solutions do not develop singularities in finite time.

Alternatively we can consider the complex Pais-Uhlenbeck oscillator where the first degree of freedom is real but the second degree of freedom has an imaginary coordinate. The correspondence principle dictates that the second degree of freedom has to be anti-Hermitian,
\begin{align}\label{eq:iPU_qp}
	q_2 = \frac{1}{\sqrt{2\omega_{i}}}(a_{2} + b_{2}) = -q_2^{\dagger},
	\qquad
	p_2  = \frac{\sqrt{\omega_{i}}}{i\sqrt{2}}(a_{2} - b_{2}) = -p_2^{\dagger},
\end{align}
implying that
\begin{align}\label{eq:iPU_ab}
	b_{2} = -a_{2}^{\dagger}.
\end{align}
This condition is consistent because the dynamics guarantees that the coordinate remains imaginary (or anti-Hermitian) as the system evolves. We stress that the last statement is independent of whether the underlying theory is classical or quantum. The resulting Hamiltonian can now be recast as
\begin{align}\label{eq:iPU_H_a1a2}
	H = \frac{\omega_1}{2} \{a_{1},a^{\dagger}_{1}\} + \frac{\omega_2}{2} \{b_{2},b^{\dagger}_{2}\}.
\end{align}
The commutation relations \eqref{eq:qPU_ab} together with \eqref{eq:iPU_ab} imply $[b_{2}, b^{\dagger}_{2}]  = 1$. The ground state is therefore defined by
\begin{align}
	a_{1} \ket{0} = 0, \qquad b_{2} \ket{0} = 0,
\end{align}
and, as before, other eigenstates can be obtained by the usual procedure of applying the creation operators $a^{\dagger}_{1}$, $b^{\dagger}_{2}$. The Hamiltonian \eqref{eq:iPU_H_a1a2} is written as a sum of commuting positive operators and is therefore positive itself. This is in perfect agreement with the correspondence principle as the energy in both the classical and the quantum case is now real and positive. 

As before, unitarity is preserved. Nevertheless, it is still highly nontrivial how to consistently include interactions because the condition \eqref{eq:iPU_qp} will generally cease to be consistent with time evolution if interactions are involved. As a result, the ghost might evolve a non-vanishing real part and render the system unstable. Moreover, the interaction term itself can be complex (or non-Hermitian) thus directly violating unitarity of time evolution.

It is also instructive to study how this relates to the Ostrogradsky instability. In section~\ref{sec:2.2}, we found  that  the generic Hamiltonian \eqref{H_second_der2} is linear in the momentum $p$ that appears only through the term $-pP$. If the variables span the entire real line, then the energy is clearly unbounded. On the other hand, the identification $p = -P^{\dagger}$ would trivially circumvent the conclusion of Ostrogradsky. This condition is usually more complicated though. Remarkably, by using the transformation rules \eqref{eq:cPU_qp12} and $Q \equiv -\ddot{q}$ the reality condition  \eqref{eq:iPU_qp} can be recast as
\begin{align}
	{\rm Re}(\ddot{q} + \omega_{1}^{2} q) = 0,	\qquad
	{\rm Im}(\ddot{q} + \omega_{2}^{2} q) = 0.
\end{align}
It is sufficient to demand this of the Cauchy data only, yet if this is required to hold at all times, then the condition \eqref{eq:iPU_qp} alone can be used to obtain the general solution \eqref{eq:PU_sol} without needing any additional input from the equations of motion \eqref{eq:eom_PU}. For $\omega_{1} \neq \omega_{2}$ the conditions \eqref{eq:PU_sol} evaluate to a real mode with frequency $\omega_{1}$ and an imaginary would-be ghost mode with frequency $\omega_{2}$.\footnote{This is not exceptional as an analogous result can also be drawn from Eq.~\eqref{eq:rPU_qp} for the real Pais-Uhlenbeck oscillator.} 

To summarise, the central message of this section is that a \emph{complex} higher derivative system with the energy bounded from below can be perfectly consistent with the Ostrogradsky theorem in both the quantum and the classical setting.

\subsection{Relation to non-canonical quantisation: negative norm states}
\label{sec:NN}

Different non-canonical quantisation schemes have been proposed that result in a bounded Hamiltonian for the Pais-Uhlenbeck oscillator~\cite{Lee:1969fy,Mannheim:2004qz,Bender:2008gh,Bender:2007wu}.  One might argue that $b_2$ should be an annihilation operator instead of a creation operator since, according to \eqref{eq:cPU_sol_ab}, it corresponds to a negative frequency. It is well known that this choice leads to negative norm states~\cite{Pauli:1943}, as we will also demonstrate shortly. The primary goal of this section is to show that quantisation of the Pais-Uhlenbeck oscillator with an indefinite metric is equivalent to canonical quantisation of the complex Pais-Uhlenbeck oscillator where the ghost degree of freedom is restricted to be imaginary. This equivalence holds up to a redefinition of the inner product.

As shown before, the Ostrogradsky theorem survives (canonical) quantisation and, based on the correspondence principle, it has been argued that is should always do so~\cite{Woodard:2015zca}. This rises the question: what is the classical limit of the non-canonically quantised higher derivative system whose energy is positive? We conjecture that it is a theory with complex variables in general and show explicitly that in the specific case of the Pais-Uhlenbeck oscillator it coincides with the system \eqref{eq:iPU_H_a1a2} containing the imaginary ghost mode. 

As we later wish to distinguish between different norms, we denote the indefinite inner product by $\braket{}{}_{\conj}$ and the corresponding conjugation by $\conj$. Also, since we focus on the negative energy degree of freedom only, the subindex 2 will be suppressed in the following to reduce clutter in notation. The reality conditions for the non-canonical choice are now expressed by $a = b^{\conj}$. Based on the previous discussion and the commutation relation~\eqref{eq:qPU_ab}, the negative norm quantisation starts by postulating 
\begin{align} \label{eq:neg_norm}
	[b,b^{\conj}] = -1, \qquad 
	b \ket{0} = 0,	\qquad
	\braket{0}{0}_{\conj} = 1.
\end{align}
This directly implies that the state $b^{\conj} \ket{0}$ has a negative norm,
\begin{align}
	\bra{0}  b b^{\conj} \ket{0}_{\conj}  = \bra{0}  [b, b^{\conj}] \ket{0}_{\conj} = -1.
\end{align}
The state space is therefore not a Hilbert space and many useful theorems break down, \eg $A = A^{\conj}$ does not imply real eigenvalues. Nevertheless, $A = A^{\conj}$ still implies that the expectation values $\bra{\psi}A\ket{\psi}_{\conj}$ are real. This leads to an inconsistent interpretation since the eigenvalues of operators might not correspond to the expectation values, \eg obsevables with an imaginary spectrum will have real expectation values.  As a specific example consider the Hamiltonian
\begin{align}\label{eq:H_ghost}
	H  = - \omega b^{\conj} b ,
\end{align}
that, by the usual arguments~\cite{Pauli:1943}, has now a positive spectrum with $E_n = \omega n$, $n \in \mathbb{Z}$. The eigenstates
\begin{align}
	\ket{n}  = \frac{(b^{\conj})^{n}}{\sqrt{n!}}\ket{0}
\end{align}
have a $\conj$ norm $-1$ for $n$ odd and a norm $+1$ for $n$ even implying that
\begin{align}
	\bra{n}H\ket{n} = (-1)^{n} E_n.
\end{align}

With negative norm states the usual definition of transition probabilities,
\begin{align}
	P(\ket{\psi} \to \ket{\phi}) = \frac{|\braket{\psi}{\phi}_{\conj}|^{2}}{\braket{\psi}{\psi}_{\conj}} ,
\end{align}
breaks down because it might predict negative probabilities. This feature is often identified with ``loss of unitarity''. An indefinite inner product of physical states is therefore in contradiction with the conventional probabilistic interpretation of quantum mechanics. It is simply not meaningful unless equipped with a consistent prescription of how to interpret the mathematics. No such interpretation is known to date.

All of the problematic issues discussed above originate from the indefiniteness of the inner product. To remedy them it is possible to define another, positive definite inner product
\begin{align}
	\braket{\psi_1}{\psi_2} \equiv \bra{\psi_1}\eta \ket{\psi_2}_{\conj},
\end{align}
where the metric operator $\eta$, defined by
\begin{align}\label{def:eta}
	\eta \ket{n} = (-1)^{n}\ket{n},
\end{align}
flips the sign of the norm for the negative norm states. This construction is well defined only if the eigenstates $\ket{n}$ form a complete basis. The conjugation with respect to the Hilbert inner product $\braket{}{}$ will be denoted by $\dagger$ as usual. The two conjugates are related as\footnote{For general $\eta$ the identity $(A^{\dagger})^{\dagger} = A$ implies $\eta = \eta^{\conj}$.}
\begin{align}\label{eq:starconj}
	A^{\dagger} = \eta^{-1} A^{\conj} \eta.
\end{align}
The metric operator is self-adjoint and unitary $\eta = \eta^{\dagger} = \eta^{-1}$.\footnote{The $\conj$ self-adjont operators are, therefore, pseudo-Hermitian operators. As they are similar to their Hermitian self-adjoints,  their spectrum is identical to their complex conjugates, \ie the eigenvalues come in complex conjugate pairs. The reality of $\conj$ expectation values implies that the eigenvalues with non-vanishing imaginary part correspond to zero $\conj$-norm eigenstates.}  The relevant non-trivial property, that follows from the defining identity \eqref{def:eta}, is that $\eta$ anti-commutes with the creation and annihilation operators,\footnote{Using Eq.~\eqref{def:eta} we obtain: $\eta b \ket{n} = \eta \sqrt{n}\ket{n-1} = (-1)^{n-1}\sqrt{n}\ket{n-1} = (-1)^{n-1}b\ket{n}= -b\eta\ket{n}$.}
\begin{align}\label{eq:etab}
	\eta b = -b \eta .
\end{align}
The reality condition $a = b^{\conj}$ now translates into $b^{\dagger} \equiv \eta b^{\conj} \eta = -a$. The sign flip, $b^{\conj} = -b^{\dagger},$ makes it possible to tell by inspection that the Hamiltonian \eqref{eq:H_ghost}, which can now be recast as $H  = \omega b^{\dagger} b$, is positive and that the  assumptions \eqref{eq:neg_norm} are consistent from the point of view of the positively defined inner product. Another peculiar consequence is that, according to \eqref{eq:cPU_abab}, the canonical coordinates
\begin{align}
	q = \frac{1}{\sqrt{2\omega}} (b - b^{\dagger}) = -q^{\dagger}, \qquad
	p =  \frac{\sqrt{\omega}}{i\sqrt{2}} (- b - b^{\dagger}) = -p^{\dagger},
\end{align}
are anti-Hermitian with respect to the positive definite inner product and thus have imaginary eigenvalues.

We arrived to this point by using a non-canonical quantisation scheme involving state space with an indefinite inner product. An identical result could have been obtained from canonical quantisation of the classical theory \eqref{eq:Lp2_PU} with $q_2$ imaginary, \ie by imposing the relation \eqref{eq:cPU_abab} with a minus sign for the ghost variables as in Eq.~\eqref{eq:iPU_qp}. In other words, the construction above is equivalent to simply noting that the following Hamiltonian,
\begin{align}
	H = -\frac{1}{2} (p^{2} + \omega^{2} q^{2}),
\end{align}
 is positive if the canonical variables are imaginary. We stress that this statement holds both in the classical or quantum theory. 
 
To elaborate on the correspondence principle note that the classical behaviour of the quantum harmonic oscillator can be well understood in terms of coherent states. The evolution of the expectation values of the canonical variables draws a trajectory in the phase space that can be obtained from the classical Hamiltonian. In fact, it follows generally from the Ehrenfest theorem that the expectation values of coordinates follow the classical equations of motion for systems whose equations of motion are linear. These considerations imply that it is relatively straightforward to test the correspondence principle for linear systems such as the Pais-Uhlenbeck oscillator.

It was shown above that the Pais-Uhlenbeck with a positive Hamiltonian requires an imaginary ghost. The negative norm formulation, on the other hand, predicts a real Pais-Uhlenbeck with a bounded Hamiltonian which can not have a classical counterpart. The last result is not satisfactory from the perspective of the correspondence principle. Therefore, we conclude that the meaningful and well behaved higher derivative theory must be complex.

\subsection{Relation to non-canonical quantisation: $PT$-symmetry}
\label{sec:PT}

Another well-known attempt of tackling the Ostrogradsky instability via non-canonical quantisation is based on $PT$-symmetric quantum mechanics. $PT$-symmetry was proposed as an alternative to Hermiticity of physical Hamiltonians~\cite{Bender:1998ke}. By this proposal, non-Hermitian but $PT$-symmetric Hamiltonians may have real and bounded eigenvalues and, therefore, may represent physical theories. This claim can be understood by noting that the $PT$-symmetric operators are similar to their Hermitian conjugates, \ie they are pseudo-Hermitian, and therefore their spectrum is identical to its complex conjugate. Thus the eigenvalues come in complex conjugate pairs and obviously are complex in general. However, it is possible to construct several $PT$-symmetric Hamiltonians with purely real spectra~\cite{Bender:1998ke,Bender:2007nj}. A comprehensive review of pseudo-Hermitian formulation of quantum mechanics can be found in Ref.~\cite{Mostafazadeh:2008pw}.

In principle, the concept of $PT$-symmetry is general and not related to the extra degrees of freedom in higher derivative theories. However, applying the idea of $PT$-symmetry
to higher derivative theories allowed the authors of Refs.~\cite{Bender:2007wu,Bender:2008gh} to make a bold claim that the ``correct" formulation of the higher derivative theories is via  $PT$-symmetry. Indeed, $PT$-symmetry allows one to define a new Hilbert space without the need for negative energy or negative norm states. Unfortunately, this leads to a specific non-canonical quantisation of the theory which does not have straightforward interpretation. In particular, the direct classical-to-quantum correspondence, which determines the physical meaning of any observable in the quantum theory, is lost. The attempts to make sense of $PT$-symmetric theories and to formulate a new, pseudo-Hermitian correspondence principle have faced several difficulties~\cite{Mostafazadeh:2008pw,Bender:2008gh,Lee:2013wfa}.

To identify the  classical theory corresponding to the $PT$-symmetric Pais-Uhlenbeck oscillator \eqref{eq:L_PU} we will closely follow the scheme put forward in~\cite{Bender:2007wu,Bender:2008gh}. It starts by expressing the model \eqref{eq:L_PU} in terms of a $PT$-symmetric Hamiltonian
\begin{align}\label{eq:PT_HPU}
	H	=	\frac{p_{x}^{2}}{2} - i p_{y}x + \frac{1}{2}(\omega_1^{2} + \omega_2^{2})x^{2} + \frac{1}{2} \omega_1^{2}\omega_2^{2} y^{2},
\end{align}
where the canonical variables $x$, $y$, $p_{x}$, $p_{y}$ are postulated to be Hermitian with respect to the standard norm. The canonical coordinates used above are clearly different from the ones used in (\ref{eq:Lp_PU}-\ref{eq:cPU_p12}). It is also apparent that the coordinate transformation contains the imaginary unit explicitly as indicated by its appearance in the Hamiltonian. The presently relevant identifications of the coordinates used in~\cite{Bender:2007wu,Bender:2008gh} with the ones in section~\ref{sec:CPU} read 
\begin{align}\label{eq:PT_qPU}
	q = i y,	\qquad \dot{q} = x.
\end{align}
We will not yet claim that the corresponding classical coordinate $q$ is imaginary. This would be at odds with the higher derivative interpretation since the speed of an imaginary variable would be real as $x$ is real. 

Similarly to the negative norm case, also the $PT$-symmetric approach is in need of different inner products. This is because the eigenvectors of non-Hermitian operators are generally not orthogonal and this in turn leads to inconsistencies with the time evolution and statistical interpretation of the expectation values even if the eigenvalues of \eqref{eq:PT_HPU} are real. As before, the unphysical inner product will be denoted by $\braket{}{}_{\conj}$ and the corresponding conjugation by $\conj$. We stress that, unlike in the previous section, the inner product $\braket{}{}_{\conj}$ is now positive definite. According to Ref.~\cite{Bender:2008gh} the physical norm, dubbed the $PT$-norm, is found to be
\begin{align}\label{eq:norm_PT}
	\braket{\psi}{\psi} = \bra{\psi}e^{-\mathcal{Q}}\ket{\psi}_{\conj},
\end{align}
where
\begin{align}
	\mathcal{Q} =  \left(\frac{p_{x}p_{y}}{\omega_1\omega_2}  + \omega_1\omega_2\, x y \right) \ln\left(\frac{\omega_1+\omega_2}{\omega_1-\omega_2}\right).
\end{align}
The norm \eqref{eq:norm_PT} is positive definite and, as before, conjugation works as $A^{\dagger} = e^{\mathcal{Q}} A^{\conj} e^{\mathcal{-Q}}$. It can be shown that the Hamiltonian \eqref{eq:PT_HPU} satisfies $H^{\dagger} = H$ if the canonical variables are self-conjugate with respect to the $\conj$ inner product. Thus $H$ has a real spectrum in the physical Hilbert space. If $H^{\dagger} = H$ and $e^{-\mathcal{Q}}$ is a positive operator then it is always possible to use $e^{-\mathcal{Q}/2}$ to define the  Hamiltonian
\begin{align}\label{eq:PT_HPU2}
	e^{-\mathcal{Q}/2} H e^{\mathcal{Q}/2} 
	= \frac{p_{x}^{2}}{2} + \frac{p_{y}^{2}}{2\omega_{1}^{2}} + \frac{1}{2} \omega_{1}^{2}x^{2}  + \frac{1}{2} \omega_{1}^{2} \omega_{2}^{2}y^{2},
\end{align}
that is now self-adjoint with respect to the $\conj$-norm. It is evident that, if $x$, $p_{x}$, $y$, $p_{y}$ are self-adjoint with respect to the $\conj$-norm, then \eqref{eq:PT_HPU2} has a positive spectrum!

To better understand what happened here we need to consider the classical counterpart of the above procedure. To this aim we note that $e^{-\mathcal{Q}/2} H(x,\ldots)e^{\mathcal{Q}/2} =  H(e^{-\mathcal{Q}/2}xe^{\mathcal{Q}/2},\ldots)$ generates the following linear complex canonical transformation 
\begin{align}\label{eq:PU_can}
&	x \to e^{-\mathcal{Q}/2} x e^{\mathcal{Q}/2} = \frac{1}{\sqrt{\omega_{1}^{2} - \omega_{2}^{2}}} 
		\left(\omega_{1} x + \frac{i p_{y}}{\omega_{1}} \right),
	\nonumber\\
&	p_{x} \to e^{-\mathcal{Q}/2} p_{x} e^{\mathcal{Q}/2} = \frac{ -i \omega_{1} \omega_{2} }{\sqrt{\omega_{1}^{2} - \omega_{2}^{2}}} 
		\left(\omega_{2} y + \frac{ip_{x}}{\omega_{2}}  \right),
	\nonumber\\
&	y \to e^{-\mathcal{Q}/2} y e^{\mathcal{Q}/2} = \frac{1}{\sqrt{\omega_{1}^{2} - \omega_{2}^{2}}} 
		\left(\omega_{1} y + \frac{i p_{x}}{\omega_{1}} \right),
	\nonumber\\
&	p_{y} \to e^{-\mathcal{Q}/2} p_{y} e^{\mathcal{Q}/2} = \frac{-i  \omega_{1} \omega_{2} }{\sqrt{\omega_{1}^{2} - \omega_{2}^{2}}} 
		\left(\omega_{2} x + \frac{i p_{y}}{\omega_{2}} \right).
\end{align}
Notably, the canonical coordinates fail to be Hermitian with respect to the new norm,  $x^{\dagger} \neq x$, if they were self-adjoint with respect to the initial norm $x^{\conj} = x$. From a quantum mechanical perspective this corresponds to a change of representation. Specifically, because $e^{-\mathcal{Q}/2}$ is not Hermitian, it transforms non-orthogonal eigenvectors of the non-Hermitian Hamiltonian \eqref{eq:PT_HPU} to orthogonal ones. The quantum-to-classical correspondence can be easily recognised by inspecting the relation between the variable $q$, \eqref{eq:PT_qPU}, and the canonical coordinates that experienced the transformation \eqref{eq:PU_can},
\begin{align}
&	q = \frac{1}{\sqrt{\omega_{1}^{2} - \omega_{2}^{2}}} 
		\left(- \frac{p_{x}}{\omega_{1}} + i \omega_{1} y\right),
	\qquad	
	\dot{q} = \frac{1}{\sqrt{\omega_{1}^{2} - \omega_{2}^{2}}} 
		\left(\omega_{1} x + \frac{ip_{y}}{\omega_{1}}  \right),
	\qquad \ldots \,\, .
\end{align}
For real $x$ and $y$ we see that the mode connected to $x$ is real while the mode connected to $y$ is imaginary. Our conclusion thus coincides with the one in the previous section: the Pais-Uhlenbeck oscillator can be split into two independent harmonic oscillators, as in Eq.~\eqref{eq:PT_HPU2}, and it has a positive definite spectrum, if the ghost degree of freedom is imaginary.

To emphasise that the $PT$-symmetric approach differs from the standard one we stress again that the authors of Refs.~\cite{Bender:2007wu,Bender:2007nj,Bender:2008gh}
postulate the pseudo-Hermitian $PT$-symmetry to be the  fundamental one, and then study its implications. Within this approach both the multiplicity of inner products and the non-orthogonal eigenstates lead to unnecessary conceptual and mathematical intricacies. Instead, our approach assumes a higher derivative complex classical system that is quantised using rules of canonical quantisation. The resulting quantum theory is mathematically consistent, possesses the usual probabilistic interpretation of quantum physics and does not clash with the correspondence principle. The energy spectrum obtains a lower bound by the almost trivial trick of demanding the would-be ghost degree of freedom to have an imaginary coordinate. We conclude, that $PT$-symmetry, although it may lead to interesting mathematical insights, is neither necessary nor practical when quantisation is concerned. The canonical quantisation of the complexified Pais-Uhlenbeck oscillator is straightforward and, moreover, it respects the correspondence principle.

\subsection{Inclusion of interactions}
\label{sec:int}

The stabilisation procedure consisted of two steps. First we complexified the theory and, second, we restricted the variables to a subspace of the complex plane that was different from the real line. It is interesting to generalise this procedure to interacting theories.  Based on the fact that the first derivative Lagrangians, such as \eqref{ex:L_1}, can have second derivative equivalents, we show how to construct stable complex higher derivative theories. The following procedure is certainly not the most general one, yet it serves as a proof of concept that it is possible to  stabilise higher derivative theories by complexification. 

First, as a specific example, consider an interacting higher derivative theory with the Lagrangian
\begin{align}\label{ex:2DL}
	L = \frac{1}{4\lambda} \frac{\ddot{q}^{2}}{q^{2}},
\end{align}
where $\lambda > 0$ is a real parameter. This Lagrangian is a special case of the system \eqref{ex:L_2} with $f_0 = 0$, $f_1 = 0$, $f_2 = \lambda q^{2}$. From Eqs.~(\ref{ex:L_0}-\ref{ex:L_2}) we then find that \eqref{ex:2DL} is equivalent  to the first derivative Lagrangian,
\begin{align}\label{ex:1DL}
	L = \frac{1}{2}\dot{q}_{1}^{2} - \frac{1}{2}\dot{q}_{2}^{2} - \frac{1}{4} \lambda \left(q_{1}^{2} - q_{2}^{2}\right)^{2},
\end{align}
describing  a particle and a ghost with the quartic interaction.  The variables of the two formulations of the same theory are related as
\begin{align}\label{ex:q1q2}
	q_{1} = \frac{1}{\sqrt{2}}\left( q - \frac{\ddot{q}}{2 \lambda q^{2}} \right),	\qquad
	q_{2} = \frac{1}{\sqrt{2}}\left( q + \frac{\ddot{q}}{2 \lambda q^{2}} \right).
\end{align}

Before continuing with studies of this system we would like to  highlight few features of the classical system. First, the higher derivative Lagrangian \eqref{ex:2DL} is invariant under a scaling transformation $q \to \Omega q$, where $\Omega$ is the scaling factor. In the first derivative Lagrangian \eqref{ex:1DL} the scale invariance is replaced by a $SO(1,1)$ symmetry. Second, the parameter $\lambda$ plays very different roles in the two formulations. In \eqref{ex:1DL}  $\lambda$ is the coupling constant, but in the higher derivative formulation $\lambda$ was introduced just for dimensional reasons -- the fourth derivative equations of motion are independent of it. The coupling constant enters the equations of motion only through the redefinition of the variables.

It is already clear from \eqref{ex:1DL} that by treating $q_{2}$ as an imaginary variable the energy will be bounded from below. This condition is consistent because the dynamics of the system can not generate a real part for $q_{2}$ if it is purely imaginary. Thus it is sufficient to require that only the initial values of $q_{2}$ and $\dot{q}_{2}$ are imaginary.

Quantisation proceeds in the standard fashion by promoting canonical variables into operators that satisfy canonical commutation relations $[q,p,] = i = [Q,P]$. Complex conjugation is replaced by Hermitian conjugation. The existence of a vacuum is ensured by choosing $q_{2}$  and $p_{2}$ anti-Hermitian, \ie by imposing
\begin{align}\label{ex:hconj}
	q_{2}^{\dagger} = -q_{2},	\qquad
	p_{2}^{\dagger} = -p_{2}.
\end{align}
From the transformation rules \eqref{ex:q1q2}, we find that the higher derivative variable now satisfies the equation of motion of a complex scalar with a $|q|^{4}$ interaction, \ie $\ddot{q} = -2 \lambda q^{2} q^{\dagger}$. As we will demonstrate shortly, this property is more general and arises from the positivity conditions.

Before moving on notice that, as in the free theory, it is possible to find a Hermitian operator $\eta$ and define the $\conj$ conjugation Eq.~\eqref{eq:starconj},
\begin{align}
	A^{\conj} \equiv \eta^{-1} A^{\dagger} \eta,
\end{align}
such that the canonical variables are $\conj$-adjoint, \ie $\eta$ can to be chosen so that $q_{i}^{\conj} = q_{i}$, $p_{i}^{\conj} = p_{i}$. Consequently the Hamiltonian of \eqref{ex:2DL} will also be $\conj$-adjoint. From Eq.~\eqref{ex:hconj} one can see that $q_{i}^{\conj} = q_{i}$, $p_{i}^{\conj} = p_{i}$ is satisfied if $\eta$ is a parity transformation of the ghost. It acts on the coordinate basis vectors $\ket{q_{1},q_{2}}$ as
\begin{align}
	\eta \ket{q_{1},q_{2}} = \ket{q_{1},-q_{2}}.
\end{align}
It follows that $\eta$ is Hermitian, unitary, the odd-parity ghost states have negative eigenvalues and, therefore, also a negative $\conj$-norm implying that the $\conj$-norm formulation does not have a consistent probabilistic interpretation. Nevertheless, since higher derivative variable is $\conj$-adjoint, $q^{\star} = q$, the $\conj$-expectation value $\langle q \rangle_{\conj}$ will be real at all times by the arguments given in Section \ref{sec:NN}.  This scheme might give the impression that it was the real higher derivative system that was quantized, yet, as in the free case, the correspondence principle does not support this conclusion. 

It is interesting to note that the canonically quantized system above does not have unitarity issues even in the presence of an imaginary would-be ghost. The unitarity violation introduced by the negative norm states thus seems to be an artificial problem. However, note that the negative norm or, equivalently, the $\eta$ odd states could be excluded from the physical spectrum thereby resolving this issue. This is possible because ghost parity is conserved in the considered system.

To generalise the above example to a broad class of models, 
we construct a stable higher derivative theory from the Lagrangian\footnote{One could equivalently start with a kinetic term of \eqref{ex:1DL}. This form of the kinetic term was chosen for simplicity.}
\begin{align}\label{eq:1Dint_L}
	L(q,\dot{q}, Q, \dot{Q}) =  \dot{q} \dot{Q} -  V(q,Q),
\end{align}
where the variables $q$ and $Q$ are complex. The energy of \eqref{eq:1Dint_L} is clearly bounded from below, if $q=Q^{*}$ and if the resulting potential $V$ is real and bounded from below. Reality of the potential is in fact required to preserve the relation $q=Q^{*}$ as the system evolves. The use of complex variables might seem exotic, yet the model above is far from that; it is nothing but an one-dimensional field theory with a complex scalar. The one-dimensional case is studied mainly for simplicity and the arguments presented here are easily generalised to any space-time dimension.

The stability conditions for the potential can be expressed as
\begin{align}
	V(q,Q)
	=	V^{*}(Q,q), 
	\qquad
	V(q,q^{*}) > V_{0},
\end{align}
for some real $V_{0}$ and general complex values of the dynamical variables. According to the construction presented in section \ref{sec:2.1} this theory is equivalent to the higher derivative theory
\begin{align}\label{eq:HDint_L}
	L'(q,\dot{q}, \ddot{q}) =  \ddot{q} Q(q, \ddot{q}) - V(q,Q(q, \ddot{q})),
\end{align}
where the function $Q(q, \ddot{q})$ is defined implicitly by the equations of motion $\ddot{q} = -\partial V/\partial Q$. The constructed higher derivative theory is equivalent to 
a first derivative theory which is well behaved if the initial conditions satisfy $q=Q^{*}$. Therefore,  if
\begin{align}\label{real_cond}
	Q(q, \ddot{q}) = q^{*}, \qquad
	\dot{Q}(q, \ddot{q}) = \dot{q}^{*},
\end{align}
then the higher derivative theory is kinetically stable. This is an explicit example of an interacting theory, where the Ostrogradsky theorem is circumvented by restricting the unstable 8-dimensional phase space to a 4-dimensional stable subspace.

Complexification doubles the number of degrees of freedom. To take an alternative point of view we could have started with a real higher derivative theory containing 
a ghost and a 4-dimensional phase space and then ``rotated'' the fields in the complex plane to obtain a different but stable theory with a equivalent number of degrees of freedom.

As in the example above, the Hamiltonian formulation of the higher derivative theory \eqref{eq:HDint_L} is identical to its normal equivalent \eqref{eq:1Dint_L}. As canonical quantisation is based on the Hamiltonian formulation, it follows that there is no difference between the canonically quantised theories corresponding to the Lagrangians \eqref{eq:1Dint_L} and \eqref{eq:HDint_L}. Especially, if the normal quantum theory is stable, then its higher derivative equivalent will also be stable.

\section{Conclusions }
  \label{4}
  
In this work we studied how complexification of higher derivative theories can avoid the consequences of Ostrogradsky theorem. The quantisation of these theories follows the canonical procedure and, as a result, satisfies the correspondence principle. Stabilisation of the physical higher derivative degrees of freedom does not, therefore, require non-canonical quantisation schemes with negative norms or with non-standard inner products of Hilbert spaces. Instead, the key ingredient for the Ostrogradsky ghosts to be physically viable is the complex physics behind the higher derivative theory.

To demonstrate that we first laid out the general properties of higher derivative theories by considering the reduction of the number of derivatives in the Lagrangian by introducing new degrees of freedom and studying  their kinetic instability. Both these features survive quantisation.  We then considered complex classical theories and their stabilisation. Based on the example of the Pais-Uhlenbeck oscillator the correspondence between a free complex classical and quantum theory was demonstrated. This approach was compared with different non-canonical quantisation schemes and it was demonstrated that removing the instability by a negative norm or $PT$-symmetric quantisation scheme is equivalent, up to a redefinition of the norm, to the canonically quantised complex classical theory. We concluded that the non-canonical quantisation schemes mainly introduce  unnecessary and impractical complications which have obscured the true essence of imaginary ghosts for a long time without giving any new insight into the problem.

Whether the imaginary ghosts have a place in physical reality remains to be shown. As a proof of concept, we presented a generic class of interacting higher derivative models without the kinetic instability together with a concrete example. Since the imaginary ghosts likely retain the needed properties for renormalization of higher derivative gravity, this would be the first obvious class of theories to apply the proposed framework.

\vskip 0.5in
\vbox{
\noindent{{\bf Acknowledgment} } \\
\noindent  
The authors thank Alberto Salvio and  Alessandro Strumia for numerous discussions. 
This work was supported by the grant IUT23-6 and by EU through the ERDF CoE program grant TK133.
}

\appendix
\section{Appendix}
\label{app}

\subsection{Second derivative equations of motion imply first derivative Lagrangians}
\label{app:der}

We prove that any Lagrangian producing second order equations of motion is equivalent to a first order Lagrangian. The proof is an adaptation of \cite{Motohashi:2014opa} where it was shown that the general third order equations of motion correspond to at most a degenerate second order Lagrangian of the general form
\begin{align}\label{eq:L_degen}
	L = \ddot{q}_{j} f_{j}(q,\dot{q}) + g_{j}(q,\dot{q}).
\end{align}
The corresponding equations of motion read
\begin{align}
	\dddot{q}_{i} \left( \frac{\partial f_{j}}{\partial \dot{q}_{i}} - \frac{\partial f_{i}}{\partial \dot{q}_{j}} \right) + \ldots = 0,
\end{align}
where the dots denote terms containing at most second derivatives. Demanding that these terms vanish, \ie that $\partial f_{j}/\partial \dot{q}_{i} = \partial f_{i}/\partial \dot{q}_{j}$, implies, by Green's theorem, that there exits a function $F(q,\dot{q})$ such that $f_{j} = \partial F / \partial \dot{q}_{j}$. By subtracting the total derivative $\dot{F}$ from \eqref{eq:L_degen} we obtain an equivalent first order Lagrangian
\begin{align}
	L = - \dot{q}_{j} \frac{\partial F }{\partial q_{j}} + g_{j}(q,\dot{q}),
\end{align}
thus proving that, up to a total derivative, second order equations of motion correspond to first order Lagrangians.

This result remains standing if the system lives in a higher dimensional space-time. To show that consider the generalisation of the degenerate Lagrangian \eqref{eq:L_degen}, 
\begin{align}
	L = q_{j,\mu\nu} f^{\mu\nu}_{j}(q,q_{,\nu}) + g_{j}(q,q_{,\nu}).
\end{align}
The corresponding third derivative terms read
\begin{align}
	q_{i,\mu\nu\sigma} \left( \frac{\partial f^{\mu\nu}_{j}}{\partial q_{i,\sigma}} - \frac{\partial f^{\mu\nu}_{i}}{\partial q_{j,\sigma}} \right) + \ldots = 0.
\end{align}
There are two ways to proceed depending on which of the third derivatives we do not wish to see in the equations of motion:
\begin{enumerate}
	\item Demanding that \emph{all} third derivatives vanish from the equations of motion implies that $\partial f^{\mu\nu}_{j}/\partial q_{i,\sigma} = \partial f^{\mu\nu}_{i}/\partial q_{j,\sigma}$. Again, Green's theorem for $\sigma = \nu$ implies that there exits a vector field $F^{\mu} (q,q_{,\nu})$ such that $f^{\mu\nu}_{j} = \partial F^{\mu} / \partial q_{j,\nu}$. It is therefore possible to remove all higher derivative terms by partial integration, \ie by subtracting $F^{\mu}{}_{,\mu}$ from the Lagrangian.

	\item  We might also demand that only third \emph{time} derivatives vanish, \ie $\partial f^{00}_{j}/\partial q_{i,0} = \partial f^{00}_{i}/\partial q_{j,0}$. Green's theorem tells then that there exists a function $F^{0} (q,q_{,\nu})$ such that $f^{00}_{j} = \partial F^{0} / \partial q_{j,0}$. Now we construct the vector $F^{\mu} = (F^{0}, 0, \ldots, 0)$ and eliminate all higher derivative terms by subtracting the total divergence $F^{\mu}{}_{,\mu}$.
\end{enumerate}

The arguments also apply to geometric theories on curved space-times, \eg  to Horndeski gravity~\cite{Horndeski:1974wa,Deffayet:2009mn}, at least locally where the action can always be expressed in terms of tensor fields and their derivatives. The equivalent first derivative Lagrangians might not be manifestly covariant, however.

\subsection{Hamiltonian formalism for second derivative Lagrangians}
\label{app:ostr}

In this Appendix we review the Hamiltonian formulation of $L(q,\dot{q}, \ddot{q})$.  First, based on the equivalent first derivative Lagrangian \eqref{def:Lprim} which we repeat here for convenience,
\begin{align}\label{def:Lprim_app}
	L'(q,\dot{q}, Q, \dot{Q}) = L(q,\dot{q}, \lambda(q,\dot{q}, Q)) - Q \lambda(q,\dot{q}, Q) - \dot{q} \dot{Q},
\end{align}
and, second, based on the Ostrogradsky prescription. The Hamiltonian corresponding to \eqref{def:Lprim_app} is defined by
\begin{align}\label{H_second_der}
	H =	\dot{q}p +	\dot{Q}P -	L',
\end{align}
for which the canonical variables read
\begin{align}\label{def:can_var}
	q, \quad
	p \equiv	\frac{\partial L'}{\partial \dot{q}} 
		= 	\frac{\partial L}{\partial \dot{q}} - \dot{Q}
	\quad \mbox{and} \quad
	Q \equiv \frac{\partial L}{\partial \ddot{q}}, \quad
	P \equiv -\dot{q}.
\end{align}
The functional dependence of \eqref{H_second_der} on the canonical variables reads
\begin{align}\label{def:reduced_H}
	H =  - Pp +  Q \lambda(q, - P, Q) - L(q, - P, \lambda(q,-P, Q)).
\end{align}

The choice of canonical variables put forward by Ostrogradsky is
\begin{align}\label{def:Orstrogradsky_var}
	q, \quad
	p \equiv \frac{\partial L}{\partial \dot{q}} 
		-	\frac{\td}{\td t}\left(\frac{\partial L}{\partial \ddot{q}}\right), 
	\quad \mbox{and} \quad
	Q \equiv \dot{q}, \quad
	P \equiv \frac{\partial L}{\partial \ddot{q}},
\end{align}
and the Hamiltonian is
\begin{align}\label{def:Orstrogradsky_H}
	H =	\dot{q}p +	\dot{Q}P -	L.
\end{align}
As expected, the Hamiltonian is linear in the momentum $p$ because it is the only quantity that depends on the third derivative $\dddot{q}$. A direct comparison reveals that the Hamiltonians \eqref{def:reduced_H} and \eqref{def:Orstrogradsky_H} and the two sets of canonical variables \eqref{def:can_var} and \eqref{def:Orstrogradsky_var} are related by the canonical transformation $Q \to -P$, $P \to Q$.

For completness we also find the conserved quantity corresponding to time translations,
\begin{align}\label{E_second_der}
	E =
	\dot{q}\left(
			\frac{\partial L}{\partial \dot{q}} 
		-	\frac{\td}{\td t}\frac{\partial L}{\partial \ddot{q}}
		\right)
	+	\ddot{q}\frac{\partial L}{\partial \ddot{q}}
	-	L.
\end{align}
Note that $\td E / \td t = -\partial L / \partial t$. A direct comparison tells that both Hamiltonians \eqref{def:Lprim_app} and \eqref{def:Orstrogradsky_H} are identical to \eqref{E_second_der}.

\bibliographystyle{JHEP}

\bibliography{im_ghost_refs}

\end{document}